# A Laser-ARPES View of the 2D Electron Systems at LaAlO$_3$/SrTiO$_3$ and Al/SrTiO$_3$ Interfaces

*Siobhan McKeown Walker, Margherita Boselli, Emanuel A. Martínez, Stefano Gariglio, Flavio Y. Bruno* and Felix Baumberger*

S. McKeown Walker, M. Boselli, S. Gariglio, F.Y. Bruno and F. Baumberger.
Department of Quantum Matter Physics, University of Geneva, 24 Quai Ernest-Ansermet, 1211 Geneva 4, Switzerland.
E. A. Martínez and F.Y. Bruno.
GFMC, Departamento de Física de Materiales, Universidad Complutense de Madrid, 28040 Madrid, Spain.
F. Baumberger.
Swiss Light Source, Paul Scherrer Institut, CH-5232 Villigen PSI, Switzerland

E-mail: fybruno@ucm.es

Keywords: 2DEG, ARPES, electronic structure, SrTiO$_3$, Al/SrTiO$_3$, LaAlO$_3$/SrTiO$_3$

## Abstract

We have measured the electronic structure of the two-dimensional electron system (2DES) found at the Al/SrTiO$_3$ (Al/STO) and LaAlO$_3$/SrTiO$_3$ (LAO/STO) interfaces by means of laser angle resolved photoemission spectroscopy, taking advantage of the large photoelectron escape depth at low photon energy to probe these buried interfaces. We demonstrate the possibility of tuning the electronic density in Al/STO by varying the Al layer thickness and show that the electronic structure evolution is well described by self-consistent tight binding supercell calculations, but differs qualitatively from a rigid band shift model. We show that both 2DES are strongly coupled to longitudinal optical phonons, in agreement with previous reports of a polaronic ground state in similar STO based 2DESs. Tuning the electronic density in Al/STO to match that of LAO/STO and comparing both systems, we estimate that the intrinsic LAO/STO 2DES has a bare band width of $\approx$ 60 meV and a carrier density of $\approx$ 6 10$^{13}$ cm$^{-2}$.

# 1. Introduction

The varied electronic phases found in complex oxides have led to a large effort towards exploiting these properties in functional devices, initiating the field of oxide electronics. [1,2] One notable example of such rich behavior is found in the two dimensional electron system (2DES) formed at the interface between the band insulators $LaAlO_3$ (LAO) and $SrTiO_3$ (STO). [3] Many interesting physical phenomena have been observed in this oxide-based 2DES, including superconductivity,[4] an intriguing magnetic response,[5,6] and an unconventional Rashba effect.[7–9] Different devices based on this system have been demonstrated, at first by writing structures with the tip of an atomic force microscope to circumvent the inherent difficulties associated with the lithographic patterning of oxides.[10] While these where finally overcome and the efficient fabrication of field effect transistors with electron beam lithography was demonstrated,[11] the high growth temperatures of the order of ~700°C needed to achieve a high mobility 2DES in LAO/STO still pose a challenge for device fabrication.[12] The demonstration that a 2DES can be formed by depositing an Al layer at room temperature onto the surface of STO has opened new perspectives for implementing STO-based 2DESs in devices.[13] The recent observation of a very large spin-to-charge conversion effect in a device based on an Al/STO 2DES highlights the potential of this system for oxide electronics.[14] The same work also suggests that the complex band structure of the 2DES is of utmost importance for its properties and for device performance.

The electronic structure of the 2DES confined at the bare surface of STO is, by now, well studied by angle resolved photoemission spectroscopy (ARPES) in the most common crystallographic orientations.[15–20] This 2DES is formed by the introduction of oxygen vacancies that are created in the bare surface by irradiation with high-energy photons under UHV conditions.[21] The same mechanism allows for the stabilization of surface 2DES in other oxides like $KTaO_3$, $SnO_2$ and $TiO_2$,[22–26] and is different to the creation of metallic STO surface layers by Ar ion bombardment.[27,28] The deposition of aluminum on the bare surface of STO in UHV creates a 2DES in a similar fashion. In this case, oxygen vacancies are created due to an efficient redox reaction, with the Al film pumping oxygen from the substrate, while it oxidizes into insulating $AlO_x$.[13] Since only very small amounts of Al are needed to induce this Al/STO 2DES, it is also accessible to surface sensitive ARPES measurements. As expected, the electronic structure of the 2DES obtained by either method is similar since both systems appear due to the presence of oxygen vacancies at the surface of STO. Note, however, that this mechanism differs from the electronic reconstruction model based on the polar discontinuity of the LAO/STO interface, which does not require the presence of oxygen vacancies at the interface.[29]

A direct measurement of the electronic structure by ARPES of the LAO/STO interfacial 2DES is desirable but the inherent surface sensitivity of the technique makes the measurement of the system buried below ≈ 15 Å of insulating LAO challenging. Attempts using both soft X-rays and conventional VUV ARPES have revealed an electronic structure qualitatively similar to that of the bare surface 2DES, but with a much broader linewidth.[30–32] However, the large density and high occupied band width reported in some of these studies,[30,31] as well as the observation by ARPES of occupied states for a LAO layer thickness far below the critical value of 4 unit cells,[32] are hard to reconcile with electronic transport measurements. This highlights the possibility that oxygen vacancies induced by the photons used to probe the system may also act as an extrinsic source of charge for the LAO/STO interface,[32,33] as they do for the bare STO surface.[15,21] Thus in order to obtain a clear picture of the intrinsic LAO/STO system by ARPES, it would be beneficial to probe the system at low photon energy, where light-induced oxygen vacancy formation is suppressed.[21] The reduced photoelectron kinetic energy at low photon energies also increases the probing depth in ARPES, which facilitates studies of buried interfaces. Furthermore, for a direct comparison of the LAO/STO and Al/STO interface 2DES, it would be desirable to tune the electronic density in the latter system to match the carrier density at the LAO/STO interface.

Here we present a laser-ARPES study of the electronic structure of the 2DESs at the Al/STO and LAO/STO interfaces. We find that their band structure, as well as the nature of electron phonon coupling at low density, is qualitatively similar to the surface 2DES studied previously in synchrotron-based VUV-ARPES experiments. We demonstrate control of the electron density of the Al/STO system by varying the thickness of the Al layer, and directly observing the changes of the Luttinger volume of the Al/STO 2DES. From a comparison with tight-binding supercell calculations, we deduce that the density observed in our experiments on Al/STO ranges from $6 \cdot 10^{13}$ cm$^{-2}$ to $3 \cdot 10^{14}$ cm$^{-2}$, which is close to the highest density that can be achieved on the bare STO surface 2DES and approaches half an electron per unit cell.[15,29] Relating these observations with our ARPES measurements of the electronic structure of the 2DES found at the LAO/STO interface we found an intrinsic electron density of approximately $6 \cdot 10^{13}$ cm$^{-2}$, in fair agreement with transport experiments. Finally, we show that the evolution of the band structure with density deviates strongly from a rigid band shift but is well captured by tight binding supercell calculations. This study paves the way for the mapping of different transport phenomena to the electronic structure.

## 2. Results

We have deposited Al on top of $TiO_2$ terminated (001)-STO substrates, after Al deposition, all Al/STO samples were transferred in-situ to the ARPES analysis chamber. Here, we will focus on samples with Al layers deposited by thermal evaporation, since these samples show the best data quality. We have, however, verified that the 2DES can also be stabilized by room temperature sputter deposition of Al (see supporting information), a widely used method in device fabrication.[14] The LAO/STO heterostructures studied in our work consist of 4 u.c. of $LaAlO_3$ deposited by pulsed laser deposition on top of a $TiO_2$ terminated (001)-STO substrate. Details of growth conditions can be found elsewhere.[34] The LAO/STO samples were exposed to air before measurement and no further step was taken to clean the surface.

Band structure calculations were performed with BinPo, a new open-source code,[35] developed following earlier work.[9,36] All carrier densities quoted in this work are obtained from the Luttinger volume of the first 4 light and 4 heavy subbands of these calculations. We find that this represents around 98% of the total charge density of the 2DES in the calculations whereas the carrier density of the first light and heavy subband quoted in Refs.[21,37] represents typically 65% of the total charge. Further details on sample growth, ARPES measurements and calculations are given in the experimental section and supporting information.

In Figure 1 we show the ARPES Fermi surfaces and energy-momentum dispersion of the 2DESs at Al/STO interfaces. All Fermi surfaces in Figure 1 (a)-(e) are formed by multiple concentric circles which is characteristic for quantum confined bands of $d_{xy}$ orbital character at the (001) surface.[9] Performing a systematic experiment on a single substrate where the Al coverage is gradually increased we find that the Fermi surface volume, and thus the density of the 2DES, increases as the Al layer thickness is incremented. This is in agreement with the proposed mechanism whereby Al pumps oxygen from the STO surface to create an Al oxide over layer leaving behind oxygen vacancies and excess itinerant electrons.[13] The discrepancy with previous ARPES work, where increasing the Al layer thickness from 2 Å to 6 Å did not increase the density of the 2DES, can be explained by considering that the mechanism may saturate at a given thickness [13,38]. Indeed, we find that further deposition of Al on the sample shown in Fig. 1(e) does not result in an increase of the Fermi surface area. Comparison with Ref.[13] and with our sputtered films with calibrated deposition rate (see supporting information) suggests that the Al thickness at which the Al/STO 2DEG density saturates in our data is ~ 2 Å. The carrier density of $3\ 10^{14}$ $cm^{-2}$ found in this case is close to the value of half an electron per unit cell found at the $GdTiO_3$/STO interface,[39] and is similar to the value at which the density of the bare surface and Al/STO 2DESs is known to saturate, as observed previously by synchrotron-ARPES[13,21]. We note that differences in the reactivity of the bare STO surface arising from slightly different recipes for the surface preparation may lead to differences in the relationship between the Aluminium overlayer thickness and the density of mobile carriers observed.[13,38,40,41] Therefore, to avoid ambiguities, we choose to discuss the Al/STO system in terms of a quantity that we observe directly in our ARPES data, namely the Fermi wave vector $k_{F,L1}$ of the first light subband and the electronic density derived from this value.

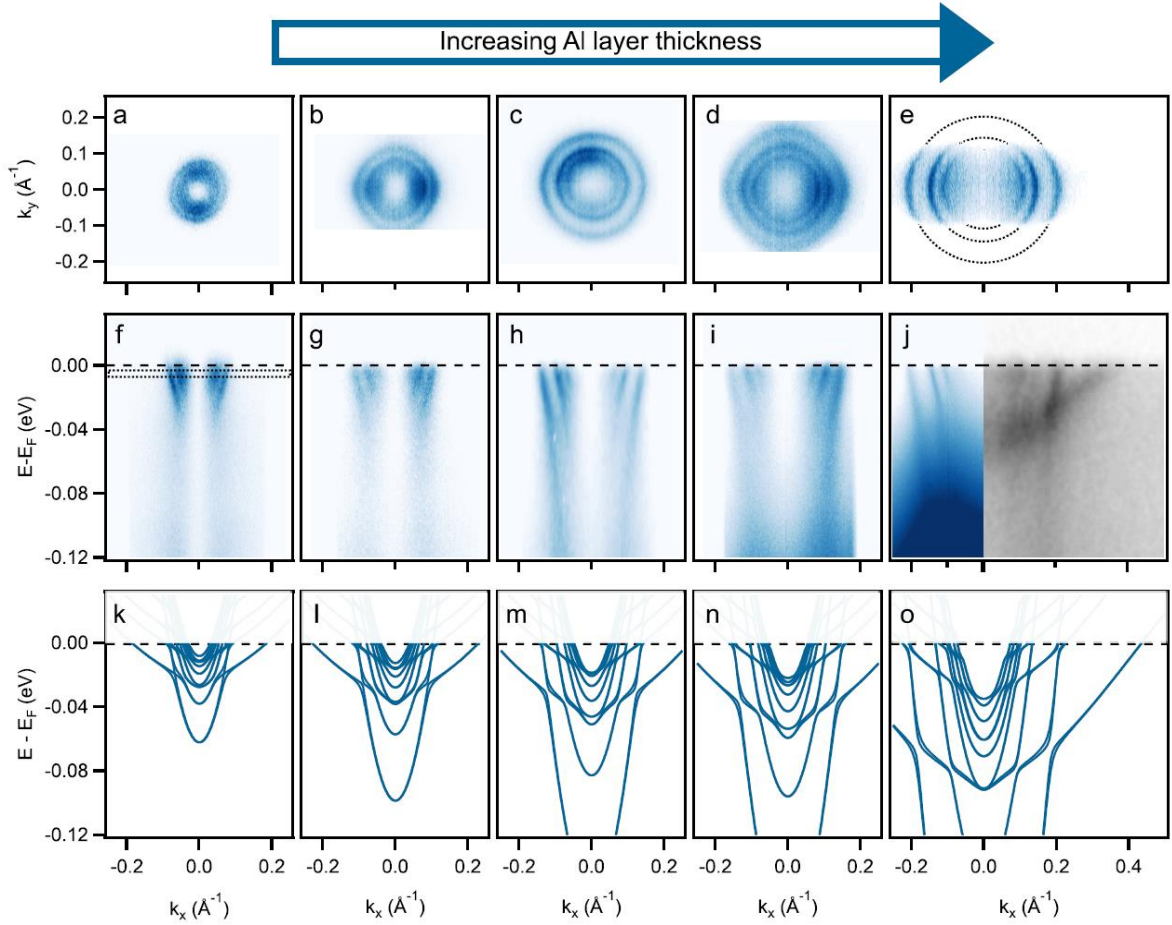

**Figure 1.** (a)-(e) ARPES Fermi surface of the two-dimensional electron system stabilized at the Al/SrTiO$_3$ interface for increasing Al thickness. (f)-(j) Energy-momentum dispersion measured along the [100] high symmetry direction ($k_y$=0). The right half of panel (j) shows synchrotron data acquired in the second Brillouin zone with a photon energy of 52 eV on a bare STO surface cleaved in UHV. All other data on Al/STO was acquired in the first Brillouin zone with hν = 6 eV. (k)-(o) Self-consistent tight binding supercell calculation of the electronic band structure. The electronic densities are from first to last column: 6, 10, 14, 17 and 31 (10$^{13}$ cm$^{-2}$).

In Figure 1(j), for $k_x > 0$, we display synchrotron data taken with light at 52 eV on a 2DES induced by UV light irradiation on a cleaved STO crystal, which has approximately the same carrier density as the Al/STO interface shown for $k_x < 0$ in the same panel. We find that the heavy $d_{xz/yz}$ states are present in the synchrotron data, but not in the laser-ARPES data.

Synchrotron measurements on Al/STO have however reported the presence of the $d_{xz/yz}$ states in Al/STO thanks to the more favourable photon energies and geometries available in those experiments [13]. Thus, we interpret the absence of the cigar shaped Fermi surfaces of the $d_{xz/yz}$ subbands in Figure 1 as a matrix element effect rather than an intrinsic property of the 2DES band structure. On the other hand, our laser ARPES data has a narrower linewidth than the best data from synchrotron experiments. This is due, in part, to the superior spectral resolution of our laser-ARPES setup but may also have a contribution from the improved homogeneity of the surfaces studied here and/or the relatively small laser spot size compared to synchrotron sources, which reduces averaging over regions of differing density. The improved linewidth of our laser ARPES data reveals kinks due to electron phonon coupling in all three visible $d_{xy}$ subbands of the most heavily doped Al/STO system, consistent with an earlier report by King et al.[9] Despite the marked improvement in spectral linewidth we cannot resolve the unconventional Rashba spin splitting of these states,[9,42] which has been predicted by a multitude of theoretical studies[8,9], and whose existence is supported by several experiments[7,14]. However, the line width of the data in Fig. 1j of 0.018 Å$^{-1}$ / 20 meV constrains the upper limit of the unconventional Rashba splitting expected at the avoided crossings of subbands to ~ 15 meV.

The band bottom of the quantum confined $d_{xy}$ states is obscured in our data due to the node in the matrix elements of $t_{2g}$ orbitals at $k_x = k_y = 0$ in this experimental geometry. We thus deduce the band width of the Al/STO 2DES by comparison with our calculations shown in Fig. 1 (k)-(o). From this we find that the bare band width of the lowest density 2DES is approximately 60 meV rising to 300 meV for the highest density 2DES. We note that the calculation does not include correlation effects and thus gives an upper limit of the quasiparticle band width.[9] Calculating the carrier density for each data set from the Fermi surface area of the first 4 subbands in our calculations, we obtain values ranging from 5.9 10$^{13}$ cm$^{-2}$ to 3.1 10$^{14}$ cm$^{-2}$.

The data in Figure 1 show that $k_{F,L1}$ varies by more than a factor of 2 from ≈ 0.085 Å$^{-1}$ to ≈ 0.21 Å$^{-1}$ with increasing Al coverage. At the same time, the splitting $\Delta k_{1,2} = k_{F,L1} - k_{F,L2}$ between the first and second light subband increases by about the same factor. This allows for a powerful test of the rigid band shift model commonly employed in the literature for the LAO/STO 2DES. To do so, we start from the same tight-binding supercell calculation shown in Figure 1(o) and rigidly shift the chemical potential to match the 5 different carrier densities determined from our data in Figure 1 (see Figure 2(a)). The values of $\Delta k_{1,2}$ found in this way are shown in Fig. 2(b) together with the experimentally observed values. This shows immediately that the trend of $\Delta k_{1,2}$ as a function of density in a rigid band shift model, is opposite to the trend observed in experiment. This behavior of the rigid band shift model is primarily a consequence of the constant (*i.e.* momentum independent) subband splitting in energy and does not depend on the details of the calculations. We thus conclude that the rigid band shift rule of thumb used to rationalize some experiments is not representative of the band structure of Al/STO 2DESs with different Al coverage.

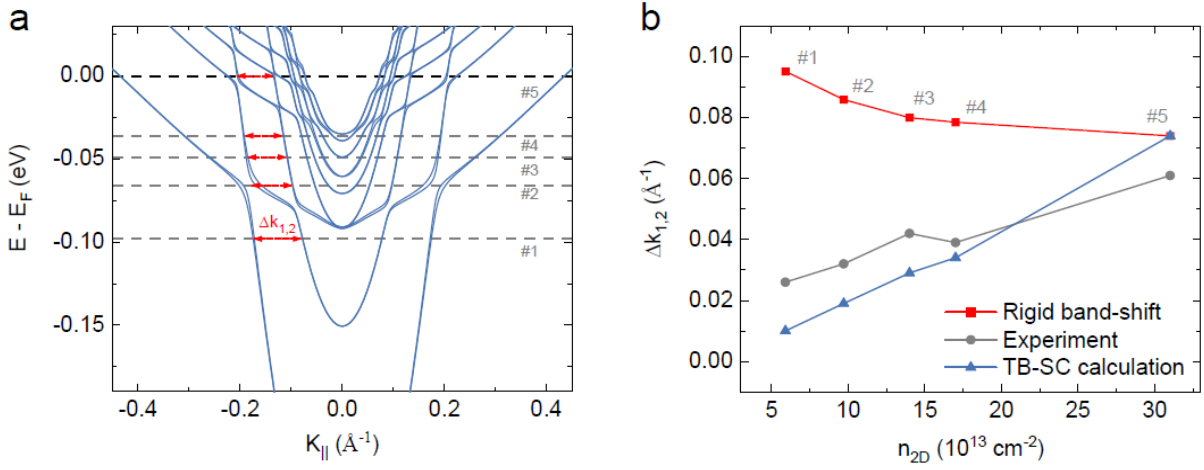

**Figure 2.** (a) Tight binding supercell calculation of the 2DES band structure for a carrier density of $3.1\ 10^{14}$ cm$^{-2}$. Dashed horizontal lines indicate the chemical potential in a rigid band shift scenario for the densities studied in our experiment: Lines labelled #5, #4, #3, #2 and #1 correspond to densities of 31, 17, 14, 10 and 6 (x $10^{13}$ cm$^{-2}$) respectively. The red arrows in (a) indicate the splitting $\Delta k_{1,2} = k_{F,L1} - k_{F,L2}$ between the first and second light subband. (b) Comparison of $\Delta k_{1,2}$ from the rigid band shift model (red) with experiment (grey) and our tight-binding supercell calculations (blue). The rigid band shift model results in values of $\Delta k_{1,2}$ that decrease with increasing density. However both the experimentally observed values of $\Delta k_{1,2}$ and those extracted from the calculations in Figure 1, increase as the 2DES density increases. Note that the rigid band shift model also results in a different relation of the Fermi wave vector $k_{F,L1}$ and density. Thus, $k_{F,L1}$ in (a) does not match the experimental value for a given density.

In contrast to the rigid band shift model, our tight-binding supercell calculations for different surface potentials shown in Figure 1 accurately reproduce the experimental trend of $\Delta k_{1,2}$. This suggests that the calculations will also provide a superior description of other band structure properties such as the unconventional spin-orbit splitting, the subband splitting's and the orbital polarization. Here we note in particular that our calculations predict that charge carriers with both heavy and light effective mass in $d_{xz/yz}$ and $d_{xy}$ bands respectively are always present in the range of densities studied here.[43]

Figure 3(a) shows the energy-momentum dispersion of the 2DES found at the LAO/STO interface measured by laser-ARPES. It is remarkable that even with the electronic system buried below 15 Å of insulating LAO, the large photoelectron escape depth at this energy allows us to probe interfacial conducting systems. A qualitative comparison with the measurements on Al/STO shown in Figure 1 reveals that the electronic structure features are well defined in energy but much broader in momentum than anticipated considering the resolution of this experiment. As a result, we cannot identify the distinct subbands that are expected to arise due to quantum confinement. This is also apparent in the Fermi surface measurements in Figure 3(b) which show a single circular blob, instead of multiple concentric contours. Similarly to the case of Al/STO, in Figure 3(a) we observe that the photoemission intensity is suppressed at $k_x = 0$, although this matrix element effect appears less pronounced in our data from LAO/STO.

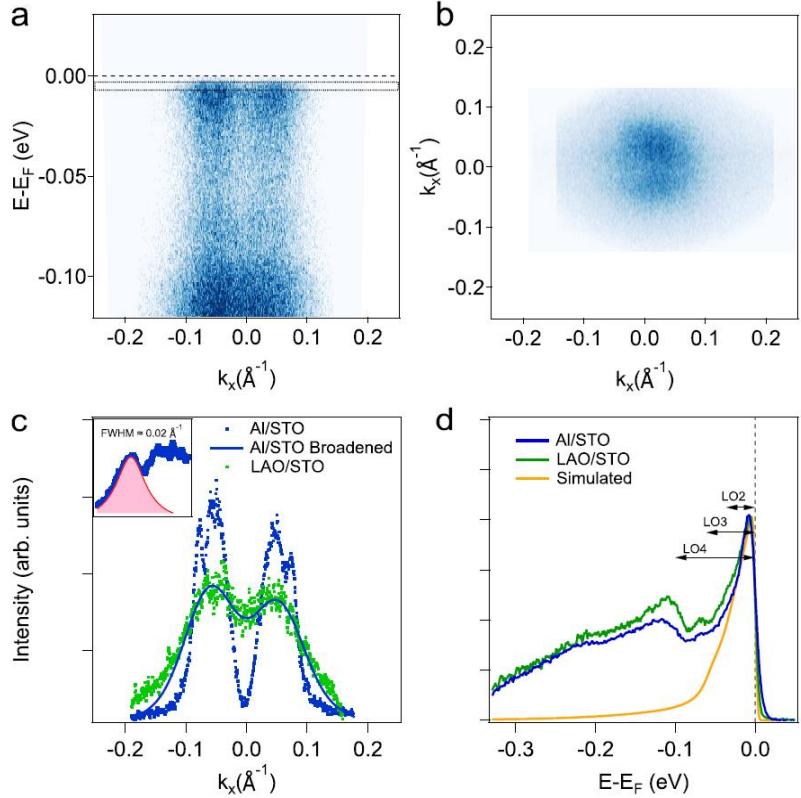

**Figure 3.** (a) Energy-momentum dispersion along the [100] high symmetry direction ($k_y=0$ plane) of the 2DES stabilized at the LAO/STO interface. (b) ARPES Fermi surface of the same electronic system. (c) Momentum distribution curves integrated over ± 2 meV around $E - E_F = 5$ meV (black rectangle in (a)) for LAO/STO (green dots) and Al/STO (blue dots). The blue solid line shows the Al/STO MDC convolved with a Gaussian with a FWHM of 0.08 Å$^{-1}$. The inset shows a Lorentzian peak fit of the first subband giving a FWHM of 0.02 Å$^{-1}$ (d) Angle-integrated energy distribution curves (EDCs) for LAO/STO (green), Al/STO (blue), and simulated spectral function (orange). The arrows indicate the energies of longitudinal optical phonon modes LO$_2$, LO$_3$ and LO$_4$ of 33 meV, 59 meV and 99 meV in bulk STO.[44,45]

In Figure 3(c) we compare momentum distribution curves (MDCs) for LAO/STO (green dots) and Al/STO at a density of 5.9 $10^{13}$ cm$^{-2}$ (blue dots). The MDCs are obtained by integrating the intensity in equivalent areas of the dispersion plots as shown by a black box in Figure 1(f) and Figure 3(a). The most striking differences between the MDCs from the two systems is that the sharp peaks originating from the first light subband in the Al/STO data, are not observed in the LAO/STO data measured with the same instrument resolution. In the inset of Figure 3(c) we show that such a peak can be fitted with a Lorentzian of full width half maximum (FWHM) of ≈ 0.02 Å$^{-1}$, which is narrow in comparison with published ARPES data, but still broader than expected from transport data of STO based 2DEGs. LAO/STO samples grown under the conditions used here have a typical low temperature mobility of $\mu$ = 300 cm$^2$ V$^{-1}$ s$^{-1}$ corresponding to an intrinsic linewidth of the order of ~ 0.003 Å$^{-1}$,[12] which is even smaller than the linewidth we observe on Al/STO. It is thus unlikely that the experimental linewidth in the LAO/STO data reflects intrinsic properties of our samples. Empirically, we find that a convolution of the Al/STO MDC with a Gaussian of FWHM 0.08 Å$^{-1}$ (shown by the solid blue line in Figure 3(c)) is remarkably similar to the MDC for LAO/STO. The momentum-integrated EDCs from both systems shown in Fig. 3(d), on the other hand, are comparable, and no additional energy broadening is observed in LAO/STO. This suggests that both electronic systems have a similar electronic density, and that there is a source of additional extrinsic momentum-broadening for the spectra of LAO/STO. Such extrinsic effects can include significant inelastic scattering of photoelectrons in the LAO overlayer, or the deflection of photoelectrons in vacuum by charge accumulated in the insulating LAO. The strong resemblance of the EDCs in the two systems indicating that there is no dramatic energy broadening in the case of LAO/STO, together with the less marked suppression of intensity at $k_x$ = 0 in LAO/STO, favours the latter scenario.

A comparison of the momentum integrated simulated spectral function (see supporting information) and energy distribution curves (EDCs) measured for both systems is shown in Figure 3(d). Both experimental EDCs present a similar structure consisting of a coherent quasiparticle peak and a main phonon satellite at approximately 100 meV forming the characteristic peak-dip-hump line shape of electrons coupled to bosonic modes. The simulated spectral function shown in orange only displays the quasiparticle peak since coupling to phonons is not taken into account in the model. Satellite peaks at ≈ 100 meV have been reported before for 2DESs at the bare STO surface and at LAO/STO interfaces and were attributed to a polaronic ground state arising from long-range coupling to the longitudinal optical phonon mode LO$_4$.[37,46,47] Here, we also identify less intense peaks at ≈ 60 meV and a faint feature at

≈ 35 meV in the EDCs presented in Figure 3(d) which have not previously been resolved in ARPES data. These energies match the zone-center frequencies of the $LO_3$ and $LO_2$ branch of bulk STO and their lower intensity is consistent with the smaller coupling constant of these modes predicted theoretically.[44,45,48] Inelastic spectral features at these energies were also reported from tunnelling experiments on LAO/STO.[49] Despite qualitative similarities between the data presented in Figure 3(d) and our previous work,[37] attempts to use the same Franck-Condon model to describe the EDCs in the present case were not satisfactory. This is primarily due to the low weight of the one-phonon satellites, which is hard to reconcile with the substantial weight in the high-energy tail of the spectra observed in our experiments. In addition, we find that the phonon satellite peaks are strongly asymmetric, which has not been reported in Refs.[37,46,47] but is predicted in theoretical work.[50]

In our earlier synchrotron work on the STO surface 2DES, we found that the Fröhlich type electron-phonon interaction is strongly suppressed at increased carrier densities where electronic screening becomes important[37]. Comparing the spectra in Figure 3(d) with data from that study for the same Fermi wave vector of $k_{F,L1} \approx 0.085$ Å$^{-1}$, and thus the same carrier density, we notice a slightly higher coherent spectral weight in the new laser-ARPES data. This suggests a certain photon energy dependence of the spectral weight, which has been proposed to indicate a final state contribution to the weight of the phonon satellites.[51] We note, however, that a recent analysis of phonon satellites in FeSe/STO over a wide photon energy range found that such final states effects are negligible at slightly higher photon energies.[52] We thus conclude that a quantitative understanding of the spectral function in STO based 2DES and related systems will require further theoretical and experimental work.

**3 Conclusion**

Our work demonstrates that the carrier density of the Al/STO 2DES - which is well suited for device integration - can be varied over a large range by controlling the Al overlayer thickness. Importantly, we find that the electronic structure measured by laser-ARPES has a non-trivial density dependence, which is well reproduced by tight-binding supercell calculations but differs qualitatively from a simple rigid band shift model. Furthermore, we find that spectra of LAO/STO measured under the same conditions show strong similarities with the data from Al/STO, including the appearance of multiple phonon replica peaks of comparable relative intensity. We thus conclude that the intrinsic electronic structure and many-body interactions in LAO/STO are very similar to the Al/STO interface at the lowest density studied here. This suggests that our band structure calculations, which quantitatively reproduce important features of the experimentally observed Al/STO band dispersion, can also provide useful insights into the LAO/STO interfacial band structure. For example, for the 2DES at the LAO/STO interface investigated here, we estimate an upper limit of the band width of 60 meV with only ≈ 20 meV splitting between the $d_{xy}$ and $d_{xz/yz}$ states and a carrier density of ≈ 6 $10^{13}$ cm$^{-2}$. This density is slightly higher than the value found in Hall-effect transport measurements performed on LAO/STO samples grown in the same conditions suggesting that not all subbands contribute equally to transport. The density estimated here also differs from ARPES data on LAO/STO measured at higher photon energy, which were likely affected much more strongly by photo-generated oxygen vacancies.[30,31] We also note that our calculations predict that charge carriers with both heavy and light effective mass in $d_{xz/yz}$ and $d_{xy}$ bands respectively are always present in the range of densities studied here. This is fully consistent with the direct observation in synchrotron data of heavy states in the STO surface 2DES down to a density ≈ 6 $10^{13}$ cm$^{-2}$ [37]. More generally, this work shows that the density dependence (and thus gating dependence) of the band structure cannot be approximated by a rigid band shift picture. It would be interesting to quantitatively compare magneto-transport data or the inverse Edelstein effect signal from spin charge conversion experiments with predictions based on self-consistent calculations such as these, to see if they might reproduce experimental observations more rigorously than the rigid band shift approach.

## 4. Experimental Section/Methods

*Al/SrTiO$_3$ samples.* We have deposited Al on top of TiO$_2$ terminated (001)-STO substrates supplied by *Crystec GmbH*, Berlin, Germany. Prior to deposition, the substrates were *in-situ* annealed for 30 min at a temperature of ~750ºC in a vacuum better than ~5 $10^{-8}$ mbar to improve surface cleanliness. The Al was deposited from independently calibrated sources in two different forms: under UHV by thermal evaporation, and by sputter deposition in a custom-built high-vacuum sputtering chamber, the details of which are described elsewhere.[53] After Al deposition, all Al/STO samples were immediately transferred in-situ to the ARPES analysis chamber. Here, we will focus on samples with Al layers deposited by thermal evaporation, since these samples show the best data quality. We have, however, verified that the 2DES can also be stabilized by room temperature sputter deposition of Al (see supporting information), a widely used method in device fabrication.[14]

*LaAlO₃/SrTiO₃ samples.* The LAO/STO heterostructures studied in our work consist of 4 u.c. of LaAlO$_3$ deposited by pulsed laser deposition on top of a TiO$_2$ terminated (001)-STO substrate supplied by *Crystec GmbH*, Berlin, Germany. Details of growth conditions can be found elsewhere.[34] The LAO/STO samples were exposed to air before measurement and no further step was taken to clean the surface. No dependence of the observed band structure on the time of exposure to the laser light was observed in LAO/STO or Al/STO, confirming that no free charges are doped into the system during the measurement process.

*ARPES measurements.* Laser-ARPES experiments were performed with a frequency converted diode laser (LEOS Solutions) providing continuous-wave radiation with 206 nm wavelength (hν = 6.01 eV) focused to a spot of ∼10 μm diameter on the sample surface and an MB Scientific electron spectrometer with a scanning lens system permitting the acquisition of two-dimensional k-space maps without rotating the sample. The measurement temperature was 5 K and the energy and momentum resolutions were 5 meV and 0.003 Å$^{-1}$, respectively.

*Band structure calculations.* Electronic structure calculations were performed with BinPo, a new open-source code.[35] In brief, we follow earlier work in Refs. [9,36] and start from a bulk density functional theory (DFT) calculation including spin-orbit coupling that is downfolded on Wannier orbitals to obtain an effective tight-binding Hamiltonian. This Hamiltonian is then diagonalized in a large supercell in the presence of a band bending potential. A self-consistent cycle is used to simultaneously satisfy the Schödinger and Poisson equations. The surface potential is chosen in the calculation in order to reproduce the Fermi wave vector $k_{F,L1}$ of the first light subband measured in our experiments, see the supporting information for a complete list of the parameters used in the calculation. All carrier densities quoted in this work are obtained from the Luttinger volume of the first 4 light and 4 heavy subbands of these calculations. We find that this represents around 98% of the total charge density of the 2DES in the calculations whereas the carrier density of the first light and heavy subband quoted in Refs.[18,37] represents typically 65% of the total charge.

**Supporting Information**
Supporting Information is available from the Wiley Online Library.

**Acknowledgements**

We acknowledge J.-M. Triscone for providing some samples for this study. This work has been supported by the Swiss National Science Foundation (Ambizione Grant No. PZ00P2-161327, project grant 165791), by Comunidad de Madrid (Atracción de Talento grant No. 2018-T1/IND-10521) and by MICINN PID2019-105238GA-I00.

Received: ((will be filled in by the editorial staff))
Revised: ((will be filled in by the editorial staff))
Published online: ((will be filled in by the editorial staff))

References

[1] M. Lorenz, M. S. Ramachandra Rao, T. Venkatesan, E. Fortunato, P. Barquinha, R. Branquinho, D. Salgueiro, R. Martins, E. Carlos, A. Liu, F. K. Shan, M. Grundmann, H. Boschker, J. Mukherjee, M. Priyadarshini, N. DasGupta, D. J. Rogers, F. H. Teherani, E. V. Sandana, P. Bove, K. Rietwyk, A. Zaban, A. Veziridis, A. Weidenkaff, M. Muralidhar, M. Murakami, S. Abel, J. Fompeyrine, J. Zuniga-Perez, R. Ramesh, N. A. Spaldin, S. Ostanin, V. Borisov, I. Mertig, V. Lazenka, G. Srinivasan, W. Prellier, M. Uchida, M. Kawasaki, R. Pentcheva, P. Gegenwart, F. Miletto Granozio, J. Fontcuberta, N. Pryds, *J. Phys. D. Appl. Phys.* **2016**, *49*, 433001.

[2] M. Coll, J. Fontcuberta, M. Althammer, M. Bibes, H. Boschker, A. Calleja, G. Cheng, M. Cuoco, R. Dittmann, B. Dkhil, I. El Baggari, M. Fanciulli, I. Fina, E. Fortunato, C. Frontera, S. Fujita, V. Garcia, S. T. B. Goennenwein, C.-G. Granqvist, J. Grollier, R. Gross, A. Hagfeldt, G. Herranz, K. Hono, E. Houwman, M. Huijben, A. Kalaboukhov, D. J. Keeble, G. Koster, L. F. Kourkoutis, J. Levy, M. Lira-Cantu, J. L. MacManus-Driscoll, J. Mannhart, R. Martins, S. Menzel, T. Mikolajick, M. Napari, M. D. Nguyen, G. Niklasson, C. Paillard, S. Panigrahi, G. Rijnders, F. Sánchez, P. Sanchis, S. Sanna, D. G. Schlom, U. Schroeder, K. M. Shen, A. Siemon, M. Spreitzer, H. Sukegawa, R. Tamayo, J. van den Brink, N. Pryds, F. M. Granozio, *Appl. Surf. Sci.* **2019**, *482*, 1.

[3] A. Ohtomo, H. Y. Hwang, *Nature* **2004**, *427*, 423.

[4] N. Reyren, S. Thiel, A. D. Caviglia, L. F. Kourkoutis, G. Hammerl, C. Richter, C. W. Schneider, T. Kopp, A.-S. Rüetschi, D. Jaccard, M. Gabay, D. A. Muller, J.-M. Triscone, J. Mannhart, *Science* **2007**, *317*, 1196.

[5] J. A. Bert, B. Kalisky, C. Bell, M. Kim, Y. Hikita, H. Y. Hwang, K. A. Moler, *Nat. Phys.* **2011**, *7*, 767.

[6] M. Salluzzo, S. Gariglio, D. Stornaiuolo, V. Sessi, S. Rusponi, C. Piamonteze, G. M. De Luca, M. Minola, D. Marré, A. Gadaleta, H. Brune, F. Nolting, N. B. Brookes, G. Ghiringhelli, *Phys. Rev. Lett.* **2013**, *111*, 087204.

[7] A. D. Caviglia, M. Gabay, S. Gariglio, N. Reyren, C. Cancellieri, J.-M. Triscone, *Phys. Rev. Lett.* **2010**, *104*, 126803.

[8] Z. Zhong, A. Tóth, K. Held, *Phys. Rev. B* **2013**, *87*, 161102.

[9] P. D. C. King, S. McKeown Walker, A. Tamai, A. de la Torre, T. Eknapakul, P. Buaphet, S.-K. Mo, W. Meevasana, M. S. Bahramy, F. Baumberger, *Nat. Commun.* **2014**, *5*, 3414.

[10] C. Cen, S. Thiel, J. Mannhart, J. Levy, *Science* **2009**, *323*, 1026.

[11] C. Woltmann, T. Harada, H. Boschker, V. Srot, P. A. van Aken, H. Klauk, J. Mannhart, *Phys. Rev. Appl.* **2015**, *4*, 064003.

[12] A. Fête, C. Cancellieri, D. Li, D. Stornaiuolo, A. D. Caviglia, S. Gariglio, J.-M. Triscone, *Appl. Phys. Lett.* **2015**, *106*, 051604.

[13] T. C. Rödel, F. Fortuna, S. Sengupta, E. Frantzeskakis, P. Le Fèvre, F. Bertran, B. Mercey, S. Matzen, G. Agnus, T. Maroutian, P. Lecoeur, A. F. Santander-Syro, *Adv. Mater.* **2016**, *28*, 1976.

[14] D. C. Vaz, P. Noël, A. Johansson, B. Göbel, F. Y. Bruno, G. Singh, S. McKeown-Walker, F. Trier, L. M. Vicente-Arche, A. Sander, S. Valencia, P. Bruneel, M. Vivek, M. Gabay, N. Bergeal, F. Baumberger, H. Okuno, A. Barthélémy, A. Fert, L. Vila, I. Mertig, J. Attané, M. Bibes, *Nat. Mater.* **2019**, *18*, 1187.

[15] W. Meevasana, P. D. C. King, R. H. He, S. Mo, M. Hashimoto, A. Tamai, P. Songsiriritthigul, F. Baumberger, Z. Shen, *Nat. Mater.* **2011**, *10*, 114.

[16] A. F. Santander-Syro, O. Copie, T. Kondo, F. Fortuna, S. Pailhès, R. Weht, X. G. Qiu, F. Bertran, A. Nicolaou, A. Taleb-Ibrahimi, P. Le Fèvre, G. Herranz, M. Bibes, N. Reyren, Y. Apertet, P. Lecoeur, A. Barthélémy, M. J. Rozenberg, *Nature* **2011**, *469*, 189.

[17] Z. Wang, Z. Zhong, X. Hao, S. Gerhold, B. Stöger, M. Schmid, J. Sánchez-Barriga, A. Varykhalov, C. Franchini, K. Held, U. Diebold, *Proc. Natl. Acad. Sci. U. S. A.* **2014**, *111*, 3933.

[18] S. McKeown Walker, A. de la Torre, F. Y. Bruno, A. Tamai, T. K. Kim, M. Hoesch, M. Shi, M. S. Bahramy, P. D. C. King, F. Baumberger, *Phys. Rev. Lett.* **2014**, *113*, 177601.

[19] S. Soltani, S. Cho, H. Ryu, G. Han, B. Kim, D. Song, T. K. Kim, M. Hoesch, C. Kim,

*Phys. Rev. B* **2017**, *95*, 125103.

[20] L. Dudy, M. Sing, P. Scheiderer, J. D. Denlinger, P. Schütz, J. Gabel, M. Buchwald, C. Schlueter, T.-L. Lee, R. Claessen, *Adv. Mater.* **2016**, *28*, 7443.

[21] S. M. Walker, F. Y. Bruno, Z. Wang, A. de la Torre, S. Riccó, A. Tamai, T. K. Kim, M. Hoesch, M. Shi, M. S. Bahramy, P. D. C. King, F. Baumberger, *Adv. Mater.* **2015**, *27*, 3894.

[22] P. D. C. King, R. H. He, T. Eknapakul, P. Buaphet, S. K. Mo, Y. Kaneko, S. Harashima, Y. Hikita, M. S. Bahramy, C. Bell, Z. Hussain, Y. Tokura, Z. X. Shen, H. Y. Hwang, F. Baumberger, W. Meevasana, *Phys. Rev. Lett.* **2012**, *108*, 1.

[23] Z. Wang, Z. Zhong, S. McKeown Walker, Z. Ristic, J.-Z. Ma, F. Y. Bruno, S. Riccò, G. Sangiovanni, G. Eres, N. C. Plumb, L. Patthey, M. Shi, J. Mesot, F. Baumberger, M. Radovic, *Nano Lett.* **2017**, *17*, 2561.

[24] F. Y. Bruno, S. McKeown Walker, S. Riccò, A. la Torre, Z. Wang, A. Tamai, T. K. Kim, M. Hoesch, M. S. Bahramy, F. Baumberger, *Adv. Electron. Mater.* **2019**, *5*, 1800860.

[25] S. Moser, L. Moreschini, J. Jaćimović, O. S. Barišić, H. Berger, A. Magrez, Y. J. Chang, K. S. Kim, A. Bostwick, E. Rotenberg, L. Forró, M. Grioni, *Phys. Rev. Lett.* **2013**, *110*, 1.

[26] J. Dai, E. Frantzeskakis, F. Fortuna, P. Lömker, R. Yukawa, M. Thees, S. Sengupta, P. Le Fèvre, F. Bertran, J. E. Rault, K. Horiba, M. Müller, H. Kumigashira, A. F. Santander-Syro, *Phys. Rev. B* **2020**, *101*, 085121.

[27] D. W. Reagor, V. Y. Butko, *Nat. Mater.* **2005**, *4*, 593.

[28] F. Y. Bruno, J. Tornos, M. Gutierrez del Olmo, G. Sanchez Santolino, N. Nemes, M. Garcia-Hernandez, B. Mendez, J. Piqueras, G. Antorrena, L. Morellón, J. De Teresa, M. Clement, E. Iborra, C. Leon, J. Santamaria, *Phys. Rev. B* **2011**, *83*, 245120.

[29] N. Nakagawa, H. Y. Hwang, D. A. Muller, *Nat. Mater.* **2006**, *5*, 204.

[30] G. Berner, M. Sing, H. Fujiwara, A. Yasui, Y. Saitoh, A. Yamasaki, Y. Nishitani, A. Sekiyama, N. Pavlenko, T. Kopp, C. Richter, J. Mannhart, S. Suga, R. Claessen, *Phys. Rev. Lett.* **2013**, *110*, 247601.

[31] C. Cancellieri, M. Reinle-Schmitt, M. Kobayashi, V. Strocov, T. Schmitt, P. Willmott, S. Gariglio, J.-M. Triscone, *Phys. Rev. Lett.* **2013**, *110*, 137601.

[32] N. C. Plumb, M. Kobayashi, M. Salluzzo, E. Razzoli, C. E. Matt, V. N. Strocov, K. J. Zhou, M. Shi, J. Mesot, T. Schmitt, L. Patthey, M. Radović, *Appl. Surf. Sci.* **2017**, *412*, 271.

[33] Z. Q. Liu, C. J. Li, W. M. Lü, X. H. Huang, Z. Huang, S. W. Zeng, X. P. Qiu, L. S. Huang, A. Annadi, J. S. Chen, J. M. D. Coey, T. Venkatesan, Ariando, *Phys. Rev. X* **2013**, *3*, 21010.

[34] M. Boselli, D. Li, W. Liu, A. Fête, S. Gariglio, J.-M. Triscone, *Appl. Phys. Lett.* **2016**, *108*, 061604.

[35] E. A. Martínez, J. I. Beltran, F. Y. Bruno, "BinPo," can be found under https://github.com/emanuelm33/BinPo, **2021**.

[36] M. S. Bahramy, P. D. C. King, a de la Torre, J. Chang, M. Shi, L. Patthey, G. Balakrishnan, P. Hofmann, R. Arita, N. Nagaosa, F. Baumberger, *Nat. Commun.* **2012**, *3*, 1159.

[37] Z. Wang, S. McKeown Walker, A. Tamai, Y. Wang, Z. Ristic, F. Y. Bruno, A. de la Torre, S. Riccò, N. C. Plumb, M. Shi, P. Hlawenka, J. Sánchez-Barriga, A. Varykhalov, T. K. Kim, M. Hoesch, P. D. C. King, W. Meevasana, U. Diebold, J. Mesot, B. Moritz, T. P. Devereaux, M. Radovic, F. Baumberger, *Nat. Mater.* **2016**, *15*, 835.

[38] L. M. Vicente-Arche, S. Mallik, M. Cosset-Cheneau, P. Noël, D. C. Vaz, F. Trier, T. A. Gosavi, C.-C. Lin, D. E. Nikonov, I. A. Young, A. Sander, A. Barthélémy, J.-P. Attané, L. Vila, M. Bibes, *Phys. Rev. Mater.* **2021**, *5*, 64005.

[39] P. Moetakef, T. A. Cain, D. G. Ouellette, J. Y. Zhang, D. O. Klenov, A. Janotti, C. G. Van de Walle, S. Rajan, S. J. Allen, S. Stemmer, *Appl. Phys. Lett.* **2011**, *99*, 232116.

[40] B. Leikert, J. Gabel, M. Schmitt, M. Stübinger, P. Scheiderer, L. Veyrat, T.-L. Lee, M. Sing, R. Claessen, *Phys. Rev. Mater.* **2021**, *5*, 065003.

[41] K. Wolff, R. Schäfer, M. Meffert, D. Gerthsen, R. Schneider, D. Fuchs, *Phys. Rev. B* **2017**, *95*, 245132.

[42] S. McKeown Walker, S. Riccò, F. Y. Bruno, A. De La Torre, A. Tamai, E. Golias, A. Varykhalov, D. Marchenko, M. Hoesch, M. S. Bahramy, P. D. C. King, J. Sánchez-Barriga, F. Baumberger, *Phys. Rev. B* **2016**, *93*, 1.

[43] A. Joshua, S. Pecker, J. Ruhman, E. Altman, S. Ilani, *Nat. Commun.* **2012**, *3*, 1129.

[44] R. A. Cowley, *Phys. Rev.* **1964**, *134*, A981.

[45] H. Vogt, *Phys. Rev. B* **1988**, *38*, 5699.

[46] C. Cancellieri, A. S. Mishchenko, U. Aschauer, A. Filippetti, C. Faber, O. S. Barisic, V. A. Rogalev, T. Schmitt, N. Nagaosa, V. N. Strocov, *Nat Commun* **2016**, *7*.

[47] C. Chen, J. Avila, E. Frantzeskakis, A. Levy, M. C. Asensio, *Nat Commun* **2015**, *6*.

[48] D. M. Eagles, *J. Phys. Chem. Solids* **1965**, *26*, 672.

[49] H. Boschker, C. Richter, E. Fillis-Tsirakis, C. W. Schneider, J. Mannhart, *Sci. Rep.* **2015**, *5*, 12309.

[50] C. A. Perroni, G. De Filippis, V. Cataudella, *Phys. Rev. B* **2021**, *103*, 245130.

[51] F. Li, G. A. Sawatzky, *Phys. Rev. Lett.* **2018**, *120*, 237001.

[52] B. D. Faeth, S. Xie, S. Yang, J. K. Kawasaki, J. N. Nelson, S. Zhang, C. Parzyck, P. Mishra, C. Li, C. Jozwiak, A. Bostwick, E. Rotenberg, D. G. Schlom, K. M. Shen, *Phys. Rev. Lett.* **2021**, *127*, 16803.

[53] E. Cappelli, W. O. Tromp, S. McKeown Walker, A. Tamai, M. Gibert, F. Baumberger, F. Y. Bruno, *APL Mater.* **2020**, *8*, 051102.

[54] O. Copie, V. Garcia, C. Bödefeld, C. Carrétéro, M. Bibes, G. Herranz, E. Jacquet, J. L. Maurice, B. Vinter, S. Fusil, K. Bouzehouane, H. Jaffrès, A. Barthélémy, *Phys. Rev. Lett.* **2009**, *102*, 1.

Supporting Information

A Laser-ARPES View of the 2D Electron Systems at LaAlO$_3$/SrTiO$_3$ and Al/SrTiO$_3$ Interfaces

*Siobhan McKeown Walker, Margherita Boselli, Emanuel A. Martínez, Stefano Gariglio, Flavio Y. Bruno\* and Felix Baumberger*

**Al/STO 2DES obtained by sputtering.**

The evaporation of Al in UHV conditions is very well suited to the study of the Al/STO 2DES by photoemission as demonstrated in the original work of Rodel *et al,* [1] and in the main body of this manuscript. However, for the fabrication of devices other growth methods such as sputtering and pulsed laser deposition are preferred.[2] We present in Figure S1 the Fermi surface of a 2DES obtained by deposition of Al by sputtering on the surface of STO with subsequent transfer of the samples to the analysis chamber under vacuum. These laser-ARPES measurements (hv = 6 eV) were obtained at a temperature of T=5K in conditions identical to those described in the main text. The commercial TiO$_2$ terminated substrates were annealed at 800°C in a pressure P≈10$^{-8}$ mbar before depositing Al under a pure P=0.18 mbar Ar atmosphere in a previously calibrated sputtering system. With increasing aluminum layer thickness the electronic density of the system increases, as shown by the increasing areas of all the Fermi contours.[2] The circular Fermi surfaces are essentially identical to those observed in the main text by Al evaporation and with the 2DES induced by photon irradiation. The possibility of tuning the 2DES density by sputtering is demonstrated, consistent with the result of Al evaporation, paving the way to the use of this technique for the fabrication of devices were the 2DES density is of fundamental importance.

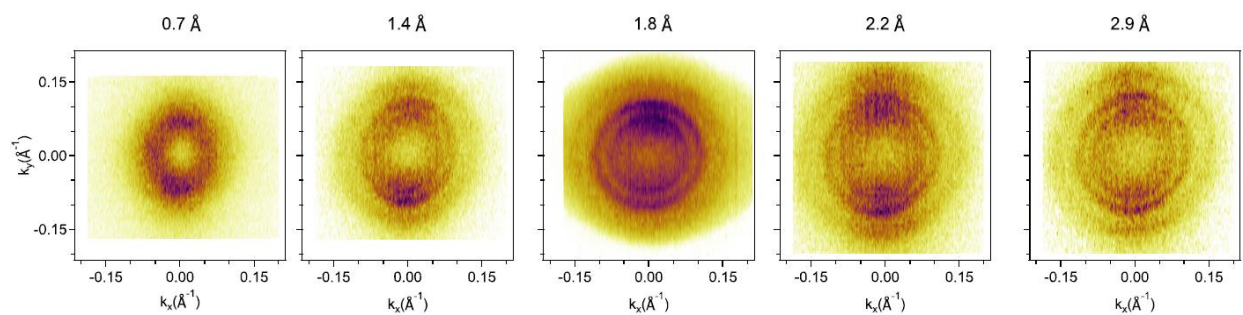

**Figure S1**. Fermi surface of the 2DEG stabilized in the (001) surface of STO after depositing Al layers of the indicated thickness by sputtering. The measurements were performed using a Laser-ARPES system with 6 eV photon energy at a temperature of T = 5 K.

**Calculations**

The calculations shown in the main text were obtained using BinPo,[3] and open source Python code. Here we give all the parameters necessary to reproduce the results shown in Figures 1(k)-(o) of the main text. The parameters of the self-consistent potential energy are the same in all the calculations except for the boundary conditions at top-most layer, $bc1$, and the constants in the relative permittivity model, $\chi_0$ and $E_c$ (see table below). The slab consisted of 50 planes stacked along the (100) direction of cubic STO. The sampling of the first Brillouin zone (BZ1) contained 46 x 46 points (with an offset of (0.001, 0.001) $Å^{-1}$). Dirichlet boundary conditions were used, setting the potential at the top-most layer to $V[0] = bc1$, while the potential at the bottom-most one was set to $V[49] = 0.0$. The shift from the lowest unoccupied level (LUL) was 0.008 $eV$ and the temperature was set to 10 $K$. The relative permittivity ($\varepsilon_r$) model used was $\varepsilon_r = 1 + \chi_0/(1 + E/E_c)$ as proposed by O. Copie et al.[4] The convergence threshold for the self-consistent potential was set to $10^{-6}$.

| Figure 1 | $bc1(eV)$ | $\chi_0$ | $E_c(V/m)$ |
|---|---|---|---|
| (k) | -0.095 | $9.5 \times 10^3$ | $3.0 \times 10^5$ |
| (l) | -0.145 | $9.5 \times 10^3$ | $4.0 \times 10^5$ |
| (m) | -0.22 | $9.0 \times 10^3$ | $5.0 \times 10^5$ |
| (n) | -0.25 | $8.0 \times 10^3$ | $7.0 \times 10^5$ |
| (o) | -0.41 | $5.0 \times 10^3$ | $2.0 \times 10^6$ |

**Table S1.** Parameters used in BinPo code for each calculation presented in Figures 1(k)-(o) of the main manuscript.

After the self-consistent solution is found, the band structure is computed using the proper post-processing component of the code. In order to reproduce the Figures 1(k)-(o), the total band structure along the XΓX path is obtained and plotted in BinPo.

**Simulated spectral function:**

We have simulated the spectral function using Lorentzian line shapes with a FWHM of 10 meV at the eigenvalues of the self-consistent tight-binding calculation. Matrix element effects are taken into account by omitting the heavy bands and by suppressing the spectral weight near $k_x=0$ in an empirical way. This is done in order to better compare with measurements. In Suplementary Figure 3a we show the results of this procedure together with the measurement of Al/STO shown in Figure 1f of the main text. The integrated spectral function convolved with a Gaussian of 5 meV FWHM to take into account the resolution of the experiment is shown in the Figure 3d of the main text.

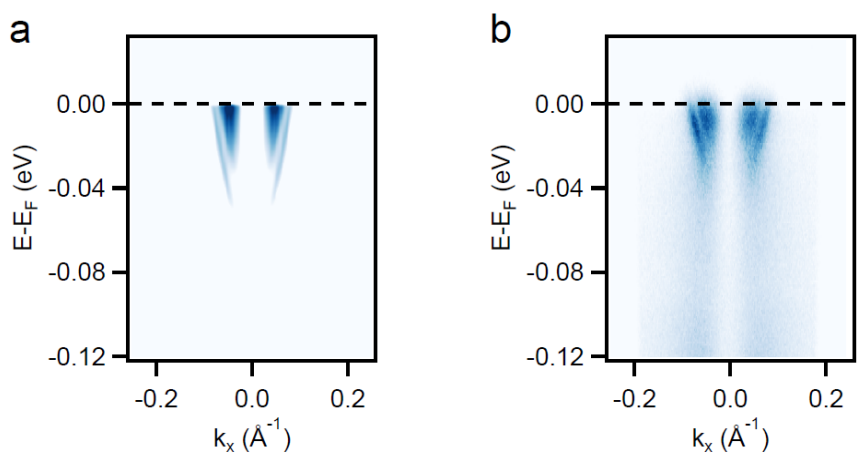

**Figure S2**. (a) Simulated spectral function using Lorentzian line shapes with a FWHM of 10 meV at the eigenvalues of the self-consistent tight-binding calculation displayed in Fig. 1k of the main text. Matrix element effects are taken into account by omitting the heavy bands and by suppressing the spectral weight near $k_x=0$ in an empirical way. (b) Energy-momentum dispersion measured along the [100] high symmetry direction in Al/STO. (This is a duplicate of Figure 1f displayed in the main text)

## References

[1] T. C. Rödel, F. Fortuna, S. Sengupta, E. Frantzeskakis, P. Le Fèvre, F. Bertran, B. Mercey, S. Matzen, G. Agnus, T. Maroutian, P. Lecoeur, A. F. Santander-Syro, *Adv. Mater.* **2016**, *28*, 1976.

[2] D. C. Vaz, P. Noël, A. Johansson, B. Göbel, F. Y. Bruno, G. Singh, S. McKeown-Walker, F. Trier, L. M. Vicente-Arche, A. Sander, S. Valencia, P. Bruneel, M. Vivek, M. Gabay, N. Bergeal, F. Baumberger, H. Okuno, A. Barthélémy, A. Fert, L. Vila, I. Mertig, J. Attané, M. Bibes, *Nat. Mater.* **2019**, *18*, 1187.

[3] E. A. Martínez, J. I. Beltran, F. Y. Bruno, "BinPo," can be found under https://github.com/emanuelm33/BinPo, **2021**.

[4] O. Copie, V. Garcia, C. Bödefeld, C. Carrétéro, M. Bibes, G. Herranz, E. Jacquet, J. L. Maurice, B. Vinter, S. Fusil, K. Bouzehouane, H. Jaffrès, A. Barthélémy, *Phys. Rev. Lett.* **2009**, *102*, 1.